\documentclass[twocolumn,preprintnumbers,amsmath,amssymb]{revtex4-1}

\usepackage{graphicx}
\usepackage{dcolumn}
\usepackage{bm}
\usepackage{multirow}
\usepackage{color}
\usepackage[normalem]{ulem}
\usepackage{amsmath}
\usepackage{gensymb}
\usepackage[colorlinks=true]{hyperref}
\usepackage{array}
\newcolumntype{P}[1]{>{\centering\arraybackslash}p{#1}}

\begin{document}

\title{Hyper-doped silicon nanoantennas and metasurfaces for tunable infrared plasmonics}
\author{Jean-Marie Poumirol$^1$}
\email{jean-marie.poumirol@cemes.fr}
\author{Cl\'ement Majorel$^1$}
\author{Nicolas Chery$^1$}
\author{Christian Girard$^1$}
\author{Peter R. Wiecha$^2$}
\author{Nicolas Mallet$^2$}
\author{Richard Monflier$^2$}
\author{Guilhem Larrieu$^2$$^,$$^4$}
\author{Filadelfo Cristiano$^2$}
\author{Anne-Sophie Royet$^3$}
\author{Pablo Acosta Alba$^3$}
\author{Sébastien Kerdiles$^3$}
\author{Vincent Paillard$^1$}
\email{vincent.paillard@cemes.fr}
\author{Caroline Bonafos$^1$}
\email{caroline.bonafos@cemes.fr}

\affiliation{\small$^1$CEMES-CNRS, Universit\'e de Toulouse, CNRS, 31055 Toulouse, France}
\affiliation{\small$^2$LAAS-CNRS, Universit\'e de Toulouse, CNRS,  31031 Toulouse, France}
\affiliation{\small$^3$CEA-LETI, Universit\'e Grenoble-Alpes, Minatec Campus, 38054 Grenoble, France}
\affiliation{\small$^4$LIMMS-CNRS/IIS, Institute of Industrial Science, The University of Tokyo, 153-8505 Tokyo, Japan.}

\begin{abstract}
\textbf{Abstract.--- } We present the experimental realization of ordered arrays of hyper-doped silicon nanodisks, which exhibit a localized surface plasmon resonance. The plasmon is widely tunable in a spectral window between 2 and 5~$\mu$m by adjusting the free carrier concentration between 10$^{20}$ and 10$^{21}$~cm$^{-3}$. We show that strong infrared light absorption can be achieved with  all-silicon plasmonic metasurfaces employing nano-structures with dimensions as low as 100~nm in diameter and 23~nm in height. Our numerical simulations show an excellent agreement with the experimental data and provide physical insights on the impact of the nanostructure shape as well as of near-field effects on the optical properties of the metasurface. Our results open highly promising perspectives for integrated all-silicon-based plasmonic devices for instance for chemical or biological sensing or for thermal imaging.

\end{abstract}


\maketitle

Common methods to enhance optical absorption and scattering, and to localize light at subwavelength dimensions are based on either plasmonic  \cite{maier_local_2003} or high refractive index dielectric nanoantennas \cite{kuznetsov_optically_2016} or metasurfaces \cite{genevet_recent_2017}.\\
In the visible spectrum, semiconductor nanostructures tend to receive a growing interest and compete nanostructures of noble metals, thanks to their low losses \cite{kuznetsov_optically_2016,boltasseva_low-loss_2011} and their direct compatibility with large-scale fabrication techniques from microelectronics.\\
In the mid-infrared (MIR), in particular at wavelengths ranging from 2 to 15~$\mu$m, region of interest for chemical or biological sensing and for thermal imaging \cite{tittl_switchable_2015, tittl_imaging-based_2018, fernandez-hurtado_enhancing_2017, bareza_mid-infrared_2020}, those approaches have limitations. For noble metals, it is difficult to support localized surface plasmon resonance (LSPR) in the MIR spectrum, because of their very high free electron concentration (about 10$^{23}$~cm$^{-3}$) \cite{luther_localized_2011}, or their behavior close to a perfect conductor. \cite{law_towards_2013}
For high index dielectric nanoantennas, the nanostructure dimensions required to send the optical Mie resonances in the MIR are huge (micron size), thus unsuitable for applications such as high resolution infrared detectors requiring a high density of small size pixels.

A promising alternative to enable plasmonics covering a large infrared spectrum are therefore doped semiconductors \cite{li_all-semiconductor_2011, baldassarre_midinfrared_2015, taliercio_semiconductor_2019}, as their LSPR is tunable through the dopant concentration in addition to other parameters such as shape and size \cite{luther_localized_2011}.
Potential candidates include impurity-doped metal oxides or vacancy-doped copper chalcogenides \cite{kriegel_plasmonic_2017}, III-V compounds \cite{law_all-semiconductor_2014}, or germanium. \cite{pellegrini_benchmarking_2018, baldassarre_midinfrared_2015, fischer_optical_2016}
Though most of these materials supply full compatibility with microelectronics fabrication processes, massively doped silicon (Si) nanostructure arrays are obviously excellent candidates for mass-production of highly integrated low-cost MIR plasmonic devices\cite{naik_alternative_2013}.

Carrier concentrations up to a few 10$^{19}$~cm$^{-3}$ are common in doped bulk Si.
Surface Plasmon Polaritons (SPP) have been demonstrated on $p$ and $n$-doped Si layers \cite{ginn_infrared_2011, shahzad_infrared_2011}, showing improved SPP confinement with respect to metal-dielectric interfaces.
However, doping Si nanostructures remains a challenge and is actually far from being mastered. 
Only few papers report on tunable LSPR in the IR range from doped Si nanostructures. 
These works involve non-thermal plasma methods and concern free-standing size-distributed nanocrystals, randomly doped at the nanocrystal level.\cite{rowe_phosphorus-doped_2013, zhou_comparative_2015, kramer_plasmonic_2015}

In this article, we show that the Si overlayer of a Silicon-On-Insulator (SOI) substrate can be massively and homogeneously doped over a wide range of active dopant concentrations, at the state of the art (from 10$^{20}$ to a few 10$^{21}$~cm$^{-3}$). We demonstrate the fabrication of a metasurface composed of dense single crystal Si nanodisks (Si-ND) on silica.
We furthermore show that such Si-ND metasurfaces strongly interact with infrared light due to a LSPR, tunable between 2.5 and 5~$\mu$m, which proves that both the concentration and scattering time of the free carriers are nearly conserved after nanostructuration process.
The experimental results are supported by numerical simulations using the Green Dyadic Method (GDM), which allowed us to explore the impact of several parameters on the LSPR, like the exact 3D-shape of the Si nanoantennas, as well as optical near-field interactions in the dense metasurface lattice.

\begin{figure}[t]
    \includegraphics*[width=\columnwidth]{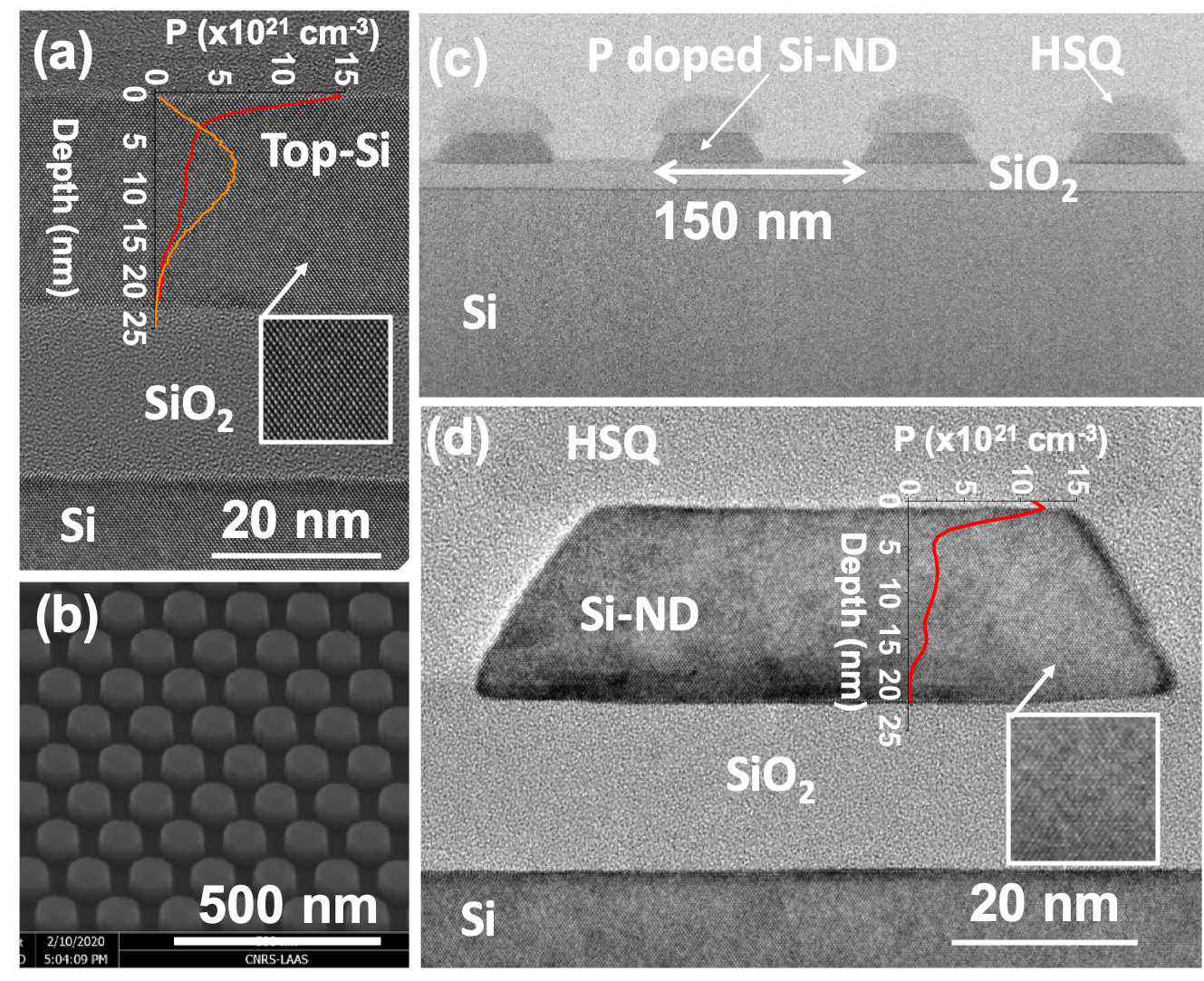}
    \caption{TEM images resuming the different fabrication steps. (a) Doping of a thin SOI layer by P implantation and LTA, illustrated by a High Resolution TEM cross-section image of the starting SOI, after implantation with P at 4~keV with a dose of 7$\times$10$^{15}$~cm$^{-2}$ and LTA at 1.45~J/cm$^2$. The as-implanted profile (orange line) as predicted by TRIM simulations \cite{ziegler_srim-2003_2004} and the STEM-EDX profile measured after LTA (red line) are superimposed; (b), (c) and (d) illustrate the second step i. e., nanopatterning of the doped layers with (b) top-view SEM image of the Si-ND hexagonal array after e-beam lithography and RIE of the hyper-doped Si overlayer; (c) Cross-section TEM bright field image of the P-doped Si-NDs; (d) Zoom in HRTEM on an individual Si-ND. Superimposed red line is the P concentration profile measured in STEM-EDX on this Si-ND. The cross-section lamellas are prepared by FIB and the NDs are cut along their diameter. In Figs.~\ref{fig:Samples}(c) and \ref{fig:Samples}(d) the apparent diameter is smaller than 100~nm. All images are taken from sample P$_3$.}
\label{fig:Samples}
  \end{figure}

Our top-down approach avoids a statistical distribution of the dopant concentration for hyper-doped Si nanostructures. In a first step, the thin Si overlayer of a SOI is  doped by means of a pulsed Laser Thermal Annealing (LTA) melt, a method which has gained recently a considerable interest in microelectronics.\cite{huet_doping_2017, cristiano_defect_2016} 
LTA allows reducing the annealing time down to a few tens of nanoseconds while still reaching the melting threshold. 
In such extremely out-of-equilibrium conditions, concentrations of electrically active dopants beyond the equilibrium solid solubility limit can be achieved in localized regions, while the integrity of the surrounding areas is preserved \cite{huet_doping_2017}.
The different steps of the elaboration process are summarized in Fig.~\ref{fig:Samples}. The starting substrates are SOI wafers with a 23~$\pm$~2~nm nominal thickness (001)-oriented Si overlayer and 20~nm thick SiO$_2$ buried oxide (BOx), lying on an intrinsic Si substrate (See Fig.~\ref{fig:Samples}(a)).

Phosphorus (P) is implanted, at low energy (4~keV), in the Si overlayer using four increasing doses (see Table~\ref{table:1}). 
For each dose, only samples corresponding to the optimal LTA conditions are considered for the next process steps, \textit{i.e.} exhibiting a defect-free single crystal Si (sc-Si) top layer.
From those layers, Si-NDs of diameter $d=100$~nm and height  $h=22$~$\pm$~1~nm as measured in Transmission Electron Microscopy (TEM) are obtained by low energy electron beam (e-beam) lithography and reactive-ion etching (RIE) \cite{y_guerfi_high_2013}. 
The Si-NDs are arranged on a dense hexagonal grid with a periodicity of $p=150$~nm, covering a square area of $50\times50$~$\mu$m$^2$ (see Scanning Electron Miceroscopy image in Fig.~\ref{fig:Samples}(b)).
Both the Si-ND diameter and lattice periodicity have been chosen to maximize the filling factor of the metasurface ($FF = 0.4$), hence maximizing the light-matter interaction for optical measurements.

Fig.~\ref{fig:Samples}(c) shows a TEM cross-section of several Si-NDs, showing that shape and dimensions are identical, and that a certain amount of under-etching causes a truncated cone shape of the disks. 
They are covered by a 20~nm Hydrogen Silesquioxane (HSQ) cap, leading to a kind of mushroom shape. The HSQ resist cap is a leftover RIE mask which we is not removed to avoid accidental lift-off of the Si-NDs.
All Si-NDs are of identical dimensions and doping level. 
The latter was evidenced by Scanning TEM - Energy Dispersive X-Ray Spectroscopy (STEM-EDX) measurement of the phosphorus redistribution upon LTA, carried out on several Si-NDs (see P concentration profile, superimposed in Fig.~\ref{fig:Samples}(d) and Fig.~S1 in Supplementary). 
A homogeneous P redistribution (plateau) is found in the melted region (Si-ND volume), however with a pile-up of P atoms at the interface of the Si-NDs with the HSQ top layer.
The P content in the plateau rises with increasing implanted dose, reaching a saturation value of 2.5$\times$10$^{21}$~cm$^{-3}$ (5 at.$\%$). 
Note that these values are identical to the ones measured in the continuous Si overlayer prior nanostructuration (see Fig.~\ref{fig:Samples}(a) and Fig.~S1 in Supplementary). 
This saturation value largely exceeds the solid solubility limit of phosphorus in bulk Si at thermal equilibrium.
The pile-up of P at the surface takes place in the annealed SOI layer before nanopatterning and is the result of liquid phase epitaxy of the silicon top layer amorphized by the P implantation. First, during melt, due to dopant diffusivities in the liquid silicon being 10$^{8}$ times greater than in solid silicon, the implanted dopants rapidly redistribute within the melted layer, so that the initial gaussian-like dopant distribution transforms into a much flatter distribution. Then, during solidification, the P atoms are pushed to the melt from solid/liquid interface, i. e. towards the surface, due to the lower solubility of P in solid Si than that in liquid Si \cite{baeri_laser_1996,reitano_solute_1994}. This mechanism implies continuous rejection of impurities on the liquid side of the moving interface until, when the total solidification is accomplished, a substantial fraction of the total amount of impurities is accumulated on the surface of the sample.

\begin{table*}[ht!]
\centering
\begin{tabular}{ |P{0.5cm}|P{2.8cm}|p{3cm}|p{3cm}|P{3cm}|P{3cm}|}
 \hline

  & Nominal implanted dose D(cm$^{-2}$)& Nominal concentration (cm$^{-3}$) & Free carrier density (cm$^{-3}$) in SOI & Scattering time $\tau$(fs) & Activation ratio ($\%$) \\ [0.5ex] 
 \hline
   \hline
P$_1$ & $1\times10^{15}$ & $9\times10^{20}$ (1.8 at.$\%$) &     $3.5\times10^{20}$ \textendash \color{blue} $3.05\times10^{20}$ &   6.15\textendash \color{blue}9.22 &  70$\pm$10\\
P$_2$ & $4\times10^{15}$ & $3.6\times10^{21}$ (7.2 at.$\%$)&   $9.2\times10^{20}$ \textendash \color{blue} $7.3\times10^{20}$ &   4.27\textendash \color{blue}5.64 &   46$\pm$10\\
P$_3$ & $7\times10^{15}$ & $6.6\times10^{21}$ (12.5 at.$\%$)& $1.4\times10^{21}$ \textendash \color{blue} $1.1\times10^{21}$ &   3.33\textendash \color{blue}4.61 &   40$\pm$9\\
P$_4$ & $1\times10^{16}$ & $9\times10^{21}$ (18at.$\%$) &   $1.6\times10^{21}$ \textendash \color{blue} $8.4\times10^{20}$  &  2.82\textendash \color{blue}3.92 &  32$\pm$15\\
\hline
\end{tabular}
\caption{Implantation conditions: nominal phosphorus implantation doses (D) at 4~keV for samples P$_1$, P$_2$, P$_3$ and P$_4$ and corresponding concentrations at the maximum of the implanted profiles, as predicted by TRIM calculations \cite{ziegler_srim-2003_2004}. Free carrier densities ($N_{\text{e}}$) and mobilities ($\mu$),  deduced from FTIR measurement on the unpatterned SOI layers. Black: average values measured on a 1~mm$^2$ surface area (data plotted in Fig.~\ref{ffig:FTIR}(b)). Blue: local values measured on a 50$\times$50$~\mu$m$^2$ at the direct vicinity of the patterned zone and used in simulations displayed in Fig.~\ref{ffig:Simu}. The average activation ratio is given by $N_{\text{e}}t/D$ with $t$ the thickness of the Si top layer where P atoms are electrically active, corresponding to the thickness (20$\pm$2~nm) amorphized by the implantation process as measured by TEM and melted during the laser annealing.}
\label{table:1}
\end{table*}

 \begin{figure}[t]
     \includegraphics*[width=8cm]{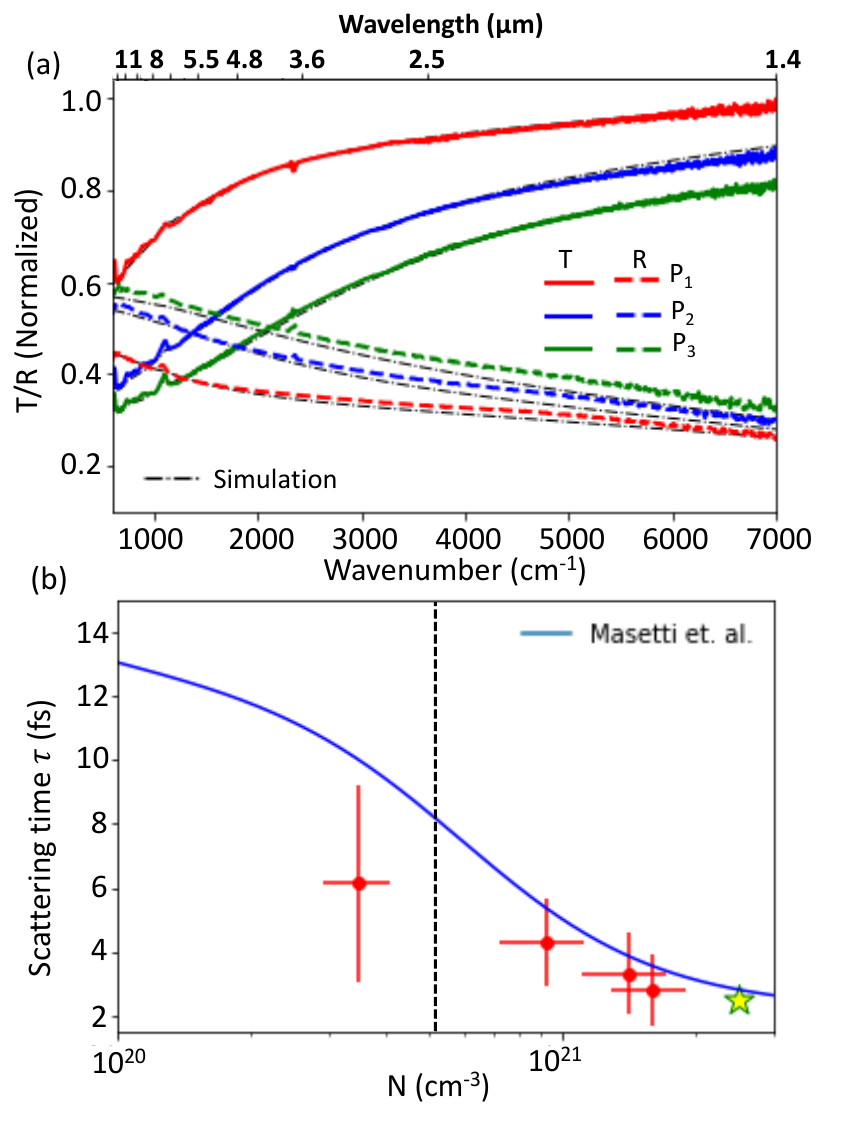}
    \caption{(a) Normalized transmittance and reflectance data. Continuous lines, measured on unpatterned hyper-doped Si layers. Dashed lines, simulated using model described in the text. (b) Scattering time versus free carrier density deduced from the fit. The blue line is the model developed in Ref.~\cite{masetti_modeling_1983} for P-doped Si. The vertical dashed line indicates the maximum electrical activation of P in Si at thermal equilibrium  \cite{pichler_intrinsic_2004}  
    and the star symbol reports the highest value measured in the literature for P activation in bulk Si, obtained after laser annealing. \cite{huet_doping_2017}}
    \label{ffig:FTIR}
  \end{figure}

Infrared spectra to quantify the free carrier absorption were measured at room temperature using a Fourier Transform Infrared Spectrometer (FTIR) coupled to a Cassegrain microscope. 
Fig.~\ref{ffig:FTIR}(a) shows the transmittance and reflectance spectra, recorded on the continuous doped layers for three implanted dopant concentrations P$_1$, P$_2$ and P$_3$ (see Table~\ref{table:1}). 
The transmittance spectra are normalized using a reference, for which the 23~nm  thick doped Si overlayer has been completely removed ($T=T_{\text{s}}/T_{\text{ref}}$), allowing the impact of this layer to be determined. 
Reflectance data are normalized to a thin area of sputtered gold on the same substrate ($R=R_{\text{s}}/R_{\text{Au}}$). 
The normalized transmittance curves display a value close to unity at high frequency, which strongly decreases at low frequency (below $\approx$ 3000~cm$^{-1}$). 
This reduced transmission is linked to light absorption by the free electrons in the $n$-doped layer. As expected, it is associated with a simultaneous increase of the reflectance. As the P concentration increases, obviously linked to an increasing free carrier density, it is evident that the transmittance minimum (resp. reflectance maximum) at low frequency is more pronounced. 
The transmittance drops to values as low as $\approx 35~\%$ for sample P$_3$ with the highest concentration.

To extract more quantitative information from our optical measurements, we model the radiative properties of the multilayer structure: doped Si ($h=23$~nm)/SiO$_2$ ($h=20$~nm)/intrinsic Si ($h=775$~$\mu$m) (Simulations in Fig.~\ref{ffig:FTIR}(a)).
The complex refractive index $n = n_1+in_2 = \sqrt{\epsilon}$ is used to calculate the complex reflectivity and transmissivity for each interface and layer. 
The radiation inside the film and inside the substrate are treated as coherent and incoherent, respectively. 
The dielectric functions for SiO$_2$ and intrinsic Si are taken from \cite{chandler-horowitz_high-accuracy_2005}. 
The dielectric function of the doped Si layer is described using a Drude-Lorentz model:
%
\begin{equation}
\epsilon_{\text{Si}} (\omega) = \epsilon_\infty^{\text{Si}} - \frac{\omega^2_{\text{p}}}{\omega(\omega+i\gamma)} \, ,
\label{feq1}
\end{equation}
%

where $\epsilon_\infty^{\text{Si}}$=13 is the value of the dielectric function at high frequency, $\omega_{\text{p}}$ = $\sqrt{4\pi N_{\text{e}}e^2/m^{*}}$ the plasma frequency, and $\gamma = 1/\tau = e/m^*\mu$ the scattering rate. 
In these expressions, $m^*= 0.3m_e$ is the electron effective mass in Si, $e$ the elementary charge, $N_{\text{e}}$ the free carrier concentration,  $\tau$ the scattering time and $\mu$ the electron mobility. \cite{kuzmenko_kramerskronig_2005}  
The free parameters used to fit the experimental data are only $N_{\text{e}}$ and  \textcolor{red} {$\tau$}, the other parameters being fixed. For a given phosphorus dose, $R$ and $T$ are fitted simultaneously \cite{kuzmenko_kramerskronig_2005}, giving one set of fitting parameters.
The dot-dashed black lines in Fig.~\ref{ffig:FTIR}(a) correspond to the best fit for each dose. The fitting lines are nearly indistinguishable from the corresponding experimental transmittance and reflectance curves for all the P doses. This excellent agreement between the experiment and a model taking into account only one Drude term suggests that the active dopant scattering rate (and therefore concentration) is mostly homogeneous (no observable doping gradient), contrary to what has been reported on thicker layers of doped silicon \cite{basu_infrared_2010}.  
This indicate that the observed phosphorus piling-up at the interface with the native SiO$_2$ shown by the EDX profile in Fig.~\ref{fig:Samples}(f) does not significantly contribute to the optical response of the material.

The extracted free carrier densities displayed in table~\ref{table:1} go from $3.5\times10^{20}$~cm$^{-3}$ for P$_1$ up to $1.6\times10^{21}$~cm$^{-3}$ for P$_4$, exceeding the solid solubility limit and most importantly the electrically active P concentration in bulk Si at thermal equilibrium \cite{pichler_intrinsic_2004} (vertical dashed line in Fig.~\ref{ffig:FTIR}(b)). 
Our best value nearly reaches the highest reported active P concentration in bulk Si obtained after pulsed laser annealing, with comparable mobilities (star symbol in Fig.~\ref{ffig:FTIR}(b)) \cite{huet_doping_2017}.
From the free carrier densities we deduced the activation ratio of implanted P ions of 70~$\%$ for P$_1$, 46~$\%$ for P$_2$, 40~$\%$ for P$_3$ and 32~$\%$ for P$_4$. One can see that not only a large fraction of P is not contributing to the free electron density, but also the activation ratio drops with increasing P concentration.

To determine the impact of the non-activated P fraction on the electronic properties, we compare in Fig.~\ref{ffig:FTIR}(b) scattering time plotted versus the free electron density $\tau(N_{\text{e}})$ to the model function developed by Masetti \textit{et al.}\cite{masetti_modeling_1983}  ($\tau_{DC} (N_{\text{e}})$) for ideal samples annealed in thermodynamical equilibrium and presenting 100~$\%$ dopant activation. 
The agreement suggests that our dopant activation process, while reaching state-of-the-art values, does not dramatically deteriorate the electron mobility and that most, if not all, of the excess P atoms form electrically inactive precipitates.\cite{nobili_precipitation_1982} 
Note that the error bars in Fig.~\ref{ffig:FTIR}(b) represent the variations of both mobility and carrier density that can be measured over the laser annealed zone. The uncertainty due to the fitting procedure is significantly smaller.

\begin{figure}[t]
	\includegraphics*[width=8.5cm]{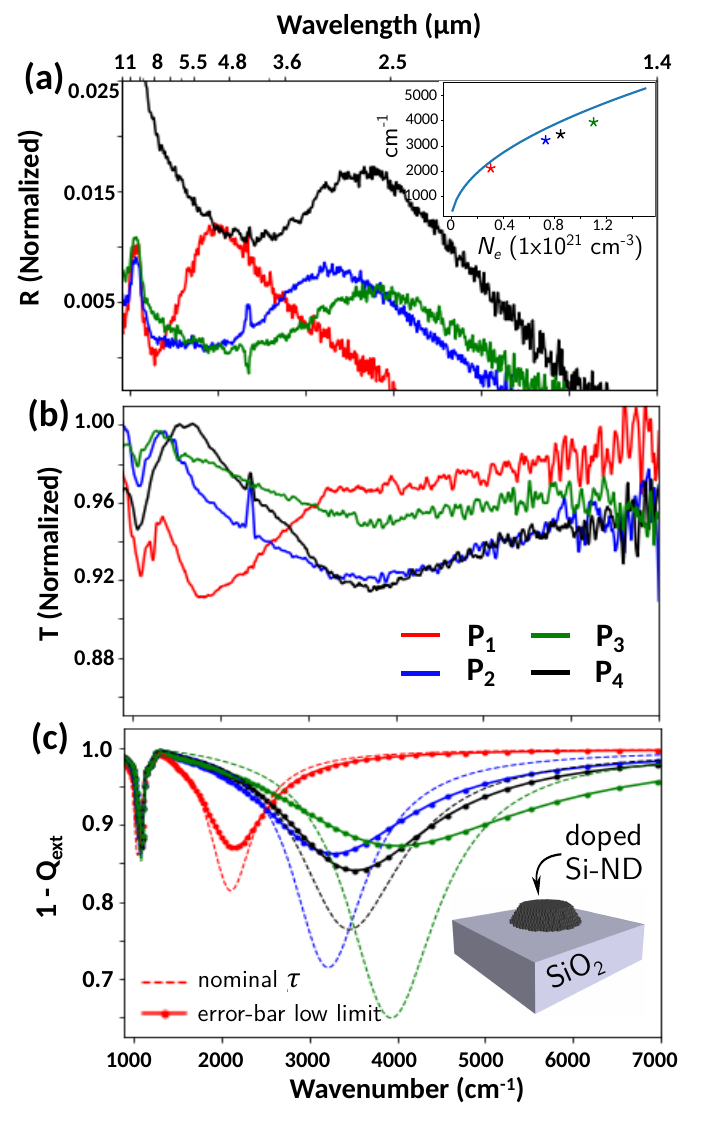} 
	\caption{(a) Normalized reflectance obtained on hyper-doped Si-ND arrays. Inset: maximum of the reflectance versus free carrier density (colored dots), Drude-Lorentz model (blue line).
		(b) Normalized transmittance obtained on hyper-doped Si-ND arrays.
		(c) Simulated extinction efficiency spectra (extinction cross-section divided by the geometrical cross-section $S=\pi (d/2)^2$ of a hyper-doped Si truncated cone lying on a SiO$_2$ substrate (see inset). Dashed lines use the mobilities as obtained by experiment, solid-dotted lines the values at their lower confidence interval. Plotted on an inverted scale to simplify qualitative comparison with the experimental transmittance spectra.
	} 
	\label{ffig:Simu}
\end{figure}

We now investigate the Si-ND metasurfaces with the four P concentrations. Fig.~\ref{ffig:Simu}(a) and (b) show their reflectance and transmittance spectra. 
The transmission minima (resp. reflection maxima) no longer appear centered at zero frequency as in the continuous layer case, but are shifted to finite frequency values. 
Furthermore, the peak shifts towards higher frequency as the free carrier density increases, unambiguously proving that a LSPR is supported by the Si-NDs. 
The variation of the plasma frequency versus $N_{\text{e}}$ is shown in inset of Fig.~\ref{ffig:Simu}(a).
It displays the typical square root close to the behavior of smaller than the wavelength particle given by:

\begin{equation}
\omega^* = \sqrt{\frac{4 \pi N_{\text{e}}e^2}{m^{*}( \epsilon_\infty^{\text{Si}}+2\epsilon_{\text{m}})}} \, ,
\end{equation}

where $\epsilon_{\text{m}}$ is the average dielectric function of the surrounding environment \cite{maier_plasmonics_2007}.

The doping range we achieved allows the Si-NDs to interact efficiently with the infrared light over a large spectral range from about 5.5 to 2.5~$\mu$m. 
For P$_1$, the LSPR is observed at 1900~cm$^{-1}$, corresponding to $\lambda_{\text{IR}} \approx 5.3$~$\mu $m. The degree of optical field confinement  achieved $\lambda_{\text{IR}}/2d = 26.5$, where $d$ is the Si-ND diameter, is comparable, at the same wavelength, to the reduction of the plasmon wavelength in graphene \cite{poumirol_magnetoplasmons_2013, tamagnone_magnetoplasmonic_2018}.
This is also far better than what can be obtained using noble metals. As a matter of fact, a micron-scale gold stripe would be necessary to support a LSPR at the same energy \cite{law_towards_2013, liu_taming_2011}. 
Thus, despite the size mismatch between wavelength and Si-NDs, at the plasmon frequency, the metasurface is able to block, either absorb or reflect, around 10~$\%$ of the incoming light.

To model our experimental results, we performed numerical simulations based on the Green Dyadic Method (GDM) with our homemade simulation toolkit ``pyGDM'' \cite{wiecha_pygdmpython_2018}, following the work done by Girard \textit{et al.} \cite{majorel_theory_2019}. 
This method is based on a volume discretization of the nanostructure and numerical solving of Maxwell's equations in the frequency domain. From the Lippmann-Schwinger equation
\begin{equation}
\begin{split}  
\mathbf{E}(\mathbf{r},\omega) = \mathbf{E_0}(\mathbf{r},\omega) + \\ 
\frac{1}{4\pi} \int_{V_{\text{j}}}\big(\epsilon_{\text{Si}}(\omega) - 1) \mathbf{S} (\mathbf{r}, \mathbf{r}', \omega). \mathbf{E}(\mathbf{r'},\omega) d\mathbf{r}'
\end{split} 
\label{equation3}
\end{equation}
the time Fourier transform of the local electric field $\mathbf{E}(\mathbf{r},\omega)$ is self-consistently obtained at any location, as a function of the incident electric field $\mathbf{E}_0(\mathbf{r}',\omega)$. The field propagator $\mathbf{S} (\mathbf{r}, \mathbf{r}',\omega)$ allows to compute the extinction and scattering efficiencies as well as the near- and far-field radiation patterns of an individual doped Si-ND.\cite{girard_shaping_2008} 

Please note that the simulations take into account the real dielectric environment (SiO$_2$ substrate), and the real shape of the antennas (a truncated cone with a base diameter of $100~$nm and top diameter of $65~$nm, as shown in inset of Fig.~\ref{ffig:Simu}(c)). 
In order to correctly describe the shape of the nanodisks, the discretization step needs to be small. 
We use 2454 dipoles on a hexagonal mesh for a Si-ND.
As a result of the reduced effective LSPR cavity size, we find that the deviation of the antenna shape from a perfect cylinder causes a 4~$\%$ redshift of the plasmon frequency.

Fig.~\ref{ffig:Simu}(c) shows the extinction efficiencies of a single Si-ND on SiO$_2$, using either the average experimental values for the scattering time (dashed lines) or their lower limits from the confidence interval (solid-dot lines, see also Fig.~\ref{ffig:FTIR}(b)).
The simulations of the extinction efficiency are in \textcolor{red}{rather good} agreement with the corresponding experimental LSPR, reproducing precisely the resonance shift as function of the free carrier concentration. 
However, while the free carrier concentration determines the spectral position of the LSPR, the width of the resonance is proportional to the damping, hence to the free-carrier scattering rate \cite{khurgin_how_2015}.
As can be seen, the lower limits for $\tau$ describe the LSPR broadening much more accurately. This might be an indication that the patterning process reduced the electron mobility in the Si-NDs, which is not  surprising as the RIE step may create defects at the Si-ND surfaces.

\begin{figure}[h]
	\includegraphics*[width=\columnwidth]{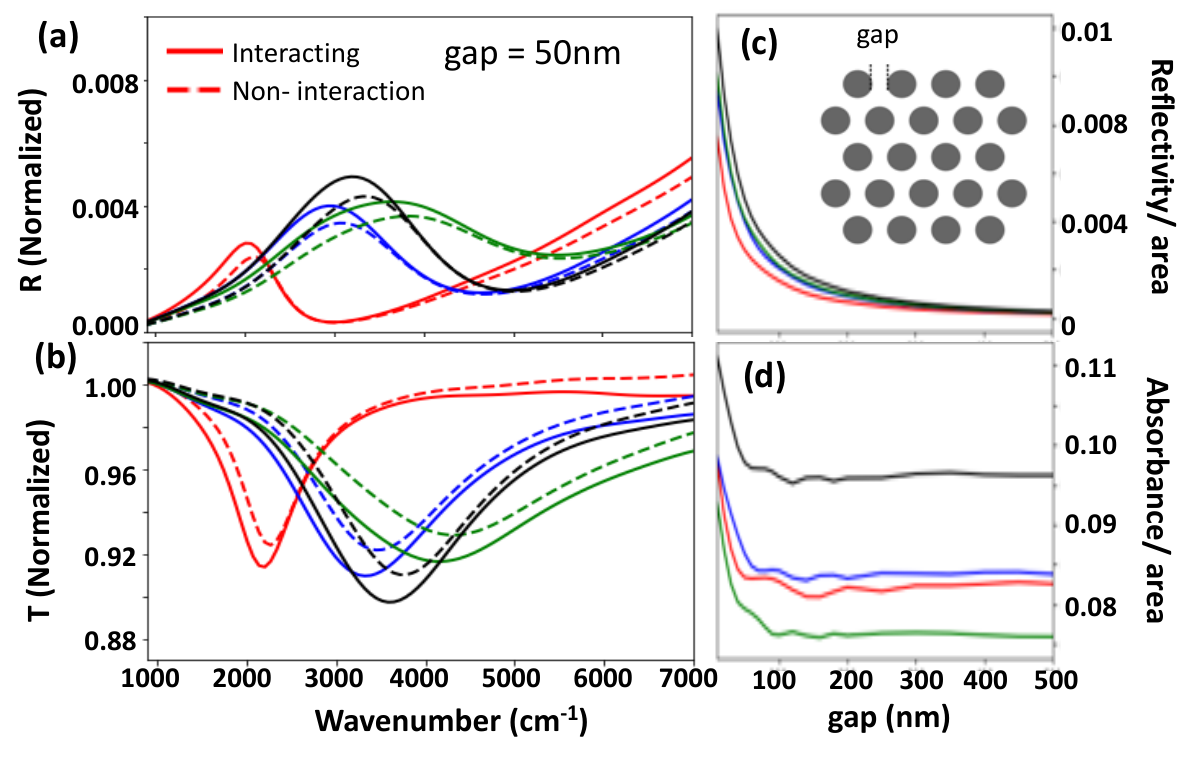}
	\caption{
		(a) Simulated reflectance of a hexagonal array metasurface with gap of 50~nm (see inset in \ref{ffig:Simu3}(c)). Simulation either with (solid lines) or without optical near-field interactions (dashed lines). A sketch of the unit cell element is depicted in the inset of Fig.~\ref{ffig:Simu}(c). 
		(b) Transmittance for the same simulations as in (a). 
		(c) Reflectance divided by the covered surface as function of the spacing (gap) between Si-NDs.
		(d) Absorptance per covered area vs. gap.
		Parameters corresponding to samples P$_1$ to P$_4$ are indicated by the same color code as in Fig.~\ref{ffig:FTIR}.
	}  
	\label{ffig:Simu3}
\end{figure}

The single-particle simulations in Fig.~\ref{ffig:Simu}(c) qualitatively reproduce the extinction, however, the according scattering efficiencies are very low (4-5 orders of magnitude lower than $Q_{\text{ext}}$, not shown).
The measured reflectance on the other hand is in the order of $0.5 - 1.5~\%$. 
To understand the emergence of the relatively pronounced reflectance, we model a hexagonal compact grid of $80\times 80$ Si-NDs, separated by a 50~nm gap between neighbors, corresponding to the nanodisk distribution on our sample (see inset in Fig.~\ref{ffig:Simu3}(c)). Such a system can \textit{per se} not be described by the GDM (too many discretization cells). 
Therefore we use a dipolar effective polarizability approximation for each Si-ND. 

The effective polarizability of the truncated cone is extracted as described by Patoux \textit{et al.}\cite{patoux_polarizabilities_2020}, and re-injected in the GDM (Eq.~\ref{equation3}), in order to calculate the optical response of the plane wave illuminated Si-ND array.
The solid lines in Fig.~\ref{ffig:Simu3}(a)-(b) show the calculated reflectance and transmittance spectra, obtained by integration of the energy flux through a plane surface above and below the metasurface and normalization to the respective energy fluxes in the empty environment (without metasurface). 
The simulations are in very good quantitative agreement with the experimental data.
To study the effect of near-field coupling and multiple scattering between the Si-NDs in the dense array, we subsequently switched off artificially the optical near-field interactions between Si-NDs in the simulation (dashed lines in Fig.~\ref{ffig:Simu3}(a)-(b)). 
We find that the transmittance (resp. the reflectance) is increased (resp. decreased), as a result of near-field coupling compared to the absence of coupling, by about 20~$\%$ near the LSPR frequency. At the same time, the resonance experiences a slight redshift.

In Fig.~\ref{ffig:Simu3}(c)-(d), we finally show the effect of the lattice periodicity (spacing between the Si-NDs) on the reflectance and absorptance at the LSPR frequency, divided by the nanodisk-covered area. The values are normalized such that the physically correct value is shown for a gap of 50~nm.
While the absorptance remains roughly constant and only increases for very small gaps $\lesssim$~50~nm, the reflectance strongly increases with decreasing spacing. 
We observe more than a 10-fold increase in the reflectance per covered area when the spacing goes from 500~nm to 50~nm. 
At the same time the normalized absorptance remains almost constant. 
Since near-field coupling causes only about $20~\%$ increase of the reflectance for a 50~nm gap (Fig.~\ref{ffig:Simu3}(a)), we attribute the over-proportional enhancement of the reflectance to constructive interference of the scattered light by the individual Si-NDs. 
This coherent effect offers an additional adjustable parameter to control the metasurface optical properties, since the fine tuning of the lattice periodicity allows to independently adjust the absorptance and reflectance (see Fig.~S3 in Supplementary for complete simulated radiation pattern).

In conclusion, by processing hyper-doped overlayers of SOI substrates, we realized all-silicon-based plasmonic metasurfaces formed by submicrometer small doped-Si nanodisks (Si-NDs). 
We demonstrated that such nanostructures support localized surface plasmon resonances, tunable via the free carrier density of the silicon. 
State-of-the-Art active dopant concentrations above 10$^{21}$~cm$^{-3}$ allow to reach the NIR regime. 
Despite a huge mismatch between the wavelength (2-5~$\mu$m) and the Si-ND diameter ($100~$nm), the LSPR enables a strong interaction with IR light resulting in a drop of the transmittance of up to 10~$\%$ at the resonance (see Fig.~S4 for performance comparaison with same size gold nanodisk). 
Via numerical simulations we investigated the optical properties of a single nanodisk as well as of the full metasurface. 
We identified collective effects and near-field coupling in the metasurface, explaining also its distinct reflectance. 
Their tunable and high absorptance renders all-silicon-based plasmonic metasurfaces very appealing for applications requiring broadband and high resolution IR detection schemes, such as micro-integrated bolometers or miniaturized IR cameras.

\indent \textbf{Acknowledgements:}~This work was partly funded by ANR DONNA (ANR-18-CE09-0034). 
The authors acknowledge the CALMIP computing facility (grant P12167), the Raimond Castaing characterization platform,the LAAS-CNRS micro and nanotechnologies platform, a member of the French RENATECH network, and SCREEN for the access to the LTA system, C. Bourgerette and S. Pinault (CEMES), F. Carcenac, B. Reig and A. Lecestre (LAAS) for helpful technical support. 

\bibliography{biblio.bib}

\end{document}